\documentstyle[prl,aps,epsf]{revtex}   
\begin{document}
\twocolumn[\hsize\textwidth\columnwidth\hsize\csname %
@twocolumnfalse\endcsname

\title{Charge Localization in Disordered Colossal-Magnetoresistance Manganites}

\author{Qiming Li,$^{1}$ Jun Zang,$^{2}$ A. R. Bishop,$^{2}$ and
C. M. Soukoulis$^{1}$ }

\address{$^1$Ames Laboratory-USDOE and Department of Physics and Astronomy,\\
 Iowa State University, Ames,
 IA 50011\\}

\address{$^2$Theoretical Division and Center for Nonlinear Studies\\ Los Alamos
National Laboratory, MS B262, Los Alamos, NM 87545 
\\}

\date{November 3, 1996}
\maketitle
\draft
\begin{abstract}

The metallic or insulating nature of the paramagnetic phase of the
colossal-magnetoresistance manganites is investigated via a double
exchange Hamiltonian with diagonal disorder. Mobility edge
trajectory is determined with the transfer matrix method.
Density of states calculations indicate that random
hopping alone is not sufficient to induce Anderson localization at the
Fermi level with 20-30\% doping. We argue that the metal-insulator
transtion is likely due to the formation of localized polarons from
nonuniform extended states as the effective band width is reduced
by random hoppings and electron-electron interactions.
\end{abstract}
\pacs{72.15.Gd, \quad 72.15.Rn,\quad  71.50.+t,\quad  71.55.+Jv, \quad  
71.38.+i}

\phantom{.} 
]
 
\narrowtext
\pagebreak

The recent discovery of colossal magnetoresistance (CMR) \cite{cmr-new}
in Mn-oxides La$_{1-x}$A$_x$MnO$_3$ (where A can be Ca, Sr, or Ba)
 has generated extensive
interests in these perovskites.  The essential correlation between 
the magnetization and resistivity \cite{cmr-old} in these compounds 
was well explained by the double exchange (DE) mechanism \cite{bib5}. 
However, the precise nature of the ferromagnetic 
metal to paramagnetic insulator transition is still not well understood.
Recently, Millis et al\cite{bib10} argued that 
Jahn-Teller (JT) effects \cite{jt-old} have to be included 
to explain the observed resistivity behavior and the magnetic
transition temperature. Millis et al \cite{millis-dm}
and Bishop and co-workers \cite{jun1} then proposed that 
the metal-insulator transition (MIT) is a consequence of the large to small 
polaron transition, induced by the reduction of effective hopping integrals
at temperatures near $T_c$.  However, there are competing 
proposals \cite{furukawa,varma} which advocate that the DE mechanism 
alone can explain the magnetic properties and MIT
in the Mn-oxides.  Notably, Varma \cite{varma} 
argued that random hopping in the paramagnetic phase due to the
DE mechanism is sufficient to localize electrons and induce a 
MIT at $T_c$.  Although there are mounting experimental evidences
of lattice effects \cite{bib12} and small polaron dynamics 
above $T_c$ \cite{polaron}, it is still not clear at the present
whether the small polaron formation for temperatures close to $T_c$ 
is the driving force of the MIT or it is merely a 
consequence of the MIT from the DE mechanism itself.
To investigate the origin of the MIT, a precise calculation 
of the localization effects in the DE model is needed.
In this Communication we study the localization properties in the 
DE model using the well-developed transfer matrix method \cite{bib20}.

The DE mechanism can be described by the following Hamiltonian:
\begin{eqnarray}
H&=& \sum_{\langle ij \rangle}
    \Bigl( t^{\gamma \gamma'}_{ij}
           c_{i \gamma \sigma}^\dagger
           c_{j \gamma' \sigma}^{\phantom{\dagger}}
                      +h.c. \Bigr) + \sum_i (\epsilon_{i\gamma}-\mu)
           n_{i \gamma}
\nonumber \\
   &-& J_H \sum_{i }
    { \vec \sigma}_{i\gamma}^{\phantom{\dagger}}
   \cdot {\vec S}_i
       +V\sum_{i}  n_{i a} n_{i b}
\ ,
\label{eq:ham}
\end{eqnarray}
where $\vec{\sigma}_{i\gamma}$ is the electronic spin
operator and $\vec{S}_i$ is the
local spin $S=3/2$ of three Mn $t_{2g}$ $d$-electrons. The operators
$c_{i\gamma\sigma}^{\phantom{\dagger}}$
($c_{i\gamma\sigma}^{\dagger}$) annihilate (create) a
mobile $e_g$ electron with spin $\sigma$ at
orbital-$\gamma$ ($\gamma\in\{a,b\}$). 
$n_{i\gamma} =c_{i\gamma\sigma}^{\dagger}c_{i\gamma\sigma}$.
The random diagonal energy $\epsilon_{i\gamma}$ is included
to account for substitutional disorders\cite{dia-dis}.
The $J_H$-term describes the Hund's rule coupling
between the local spin of $t_{2g}$ electrons and the $e_g$ electrons;
the $V$-term describe the on-site inter-orbital electron-electron (e-e)
interactions. In the Mn-oxides, 
$J_H\gg t/S$. Here we neglect the on-site intra-orbital e-e 
interactions and exchange interactions, which is reasonable in 
the large $J_H$ limit.
 
We consider only the limit 
$SJ_H/t \rightarrow\infty$. As a consequence,  the electronic spin at each
site is parallel to the local spin,  for the low energy states of interest.  
To study these low energy states, we can use a projection operator 
${\cal P}$ to project out the high energy states. Then the Hund's rule 
coupling interaction can be dropped and
 the hopping term is renormalized to 
\begin{equation}
t_{ij}=t[cos(\frac{\theta_i} {2})cos(\frac{\theta_j} {2}) +
  sin(\frac{\theta_i} {2})sin(\frac{\theta_j} {2})e^{i(\phi_i-\phi_j)}], 
\label{eq:tij}
\end{equation}
in the classical spin limit $S=\infty$, where 
$(\theta_i,\phi_i)$ is the polar angles of the classical spin
at site-$i$. Both finite $J_H$ and $S$ will decrease the
localization effect, while the inter-orbital e-e
interaction $V$ will probably increase it.
To simplify the problem further, we will neglect
the e-e interaction here and assume that its only important contribution
is to renormalize Hamiltonian parameters. The Hamiltonian  
(\ref{eq:ham})  is then transformed to an effective one-electron Hamiltonian
of the form 
\begin{equation}
H=\sum_i (\epsilon_i-\mu) n_{i} + 
\sum_{<ij>} (t_{ij}  c_{i}^\dagger c_{j}^{\phantom{\dagger}} + H.c.).
\label{eq:ham2}
\end{equation}
The Hamiltonian (\ref{eq:ham2}) describes the dynamics
of electrons in a {\it static} background of classical
spins, with the hopping matrix elements dependent on the
spin configuration via Eq.(\ref{eq:tij}).
This approximation of dynamic disorder by static disorder
is valid in the large $S$ limit. Note that we also replace
the two-orbitals by one effective orbital. If there is no
Jahn-Teller coupling, the carrier density of
each orbital would be approximately half of the
doping $x$. However, static or dynamic
JT effects will complicate this issue and
the exact correspondence is not well understood at the metallic 
region at present.

In the simplified Hamiltonian (\ref{eq:ham2}), electron localization
is due to both off-diagonal and diagonal disorder. The off-diagonal
disorder is intrinsic to the DE model in the paramagnetic
phase, and the resistivity due to this spin-disordered
scattering was calculated in Born-approximation \cite{fisher} or
memory function approximation \cite{bib10}. Localization via 
off-diagonal disorder has not been studied extensively. 
A few calculations\cite{bib19} have been done for uniform
distributions of real hopping integrals $t$ in one and two-dimensional systems.
Also there are studies on 2D random flux systems 
with uniform $|t|$ \cite{review}.
The most relevant work for the present model was a 
study by Economou and Antoniou\cite{bib15} for systems with a 
semicircle distribution of hopping integrals. Their work was based 
on the localization criteria 
of Economou and Licciardello\cite{bib16}. 
A recent calculation\cite{bib17} based 
on the Ziman criterion\cite{bib18} was also reported. 
None of these calculations
produce the precise mobility edge. Moreover, the random Berry's
phases (the phase of $t_{ij}$ in Eq.(\ref{eq:tij}))
are not included, and their importance is difficult to 
assess. 

Here we investigate the localization properties of Hamiltonian (\ref{eq:ham2}) 
with the transfer matrix method\cite{bib20}. This technique, coupled with
finite size scaling analysis, produces the most reliable information
about the extended or localized nature of the eigenstates. In this technique,
one considers a bar of length $N$ and cross section $M\times M$. One determines
the largest localization length $\lambda_M$ as $N\rightarrow \infty$ from 
the smallest Lyapunov coefficient of the product of the random transfer 
matrix relevant to Eq.(\ref{eq:ham2}). The nature of the eigenstate 
can be determined by studying the scaling property of the localization length
of finite systems.  For extended (localized) states, $\lambda_M/M$ 
increases (decreases) with increasing M. At the mobility edge, which separates
the extended from the localized states, $\lambda_M/M$ is independent of M
and this behavior defines the Anderson transition. 

In the paramagnetic phase, 
the direction of the local spin $\vec{S}_i$ is chosen to be 
uniformly distributed on a sphere: $P(\phi_i)=1/(2\pi)$;
$P(\cos(\theta_i))=1/2$. 
Once the spin configuration $\{\vec{S_i}\}$ is specified,
the nearest neighbor hopping integrals can be obtained via Eq.(\ref{eq:tij}). 
The random diagonal site-energy $\epsilon_i$ is assumed to be 
distributed uniformly  between [-W/2, W/2]. In our calculation, 
we have used systems with widths $M=4-14$ and the length N on the order 
of N = 30,000 to minimize errors.

We first investigate the localization effects in the absence of any 
diagonal disorder, W = 0, to see whether random hoppings with Berry phases
in the paramagnetic phase is alone sufficient to lead to localization
of a large fraction of electrons, as has been argued in \cite{varma}.  
Our results are presented in Fig.~1, in which we plot the ratio of the 
calculated finite size localization length $\lambda_M$ with M,  
as a function of M for different energies. It is clear that the mobility edge
E$_c$,  where $\lambda_M/M$ should be a constant independent of M, 
is located between $3.55< |E_c|/t < 3.6$. 
To determine whether the Fermi level is below or above E$_c$ at certain doping
levels, we need to know the density of states (DOS) of the system,  which we 
obtain by directly diagonalizing the Hamiltonian matrix for a finite size
(10x10x10) cluster. The integrated DOS and the DOS, 
averaged over many (100) configurations, are shown in Fig.~2. 
This indicates that less than 0.5\% of states are below $E_c =-3.56t$ (Fig.~2), 
far less than the required 20-30\% for the CMR system. Therefore, 
the present calculation confirms the suggestion \cite{millis-dm} that 
purely off-diagonal disorder in the DE model is far from sufficient to 
localize electrons in the CMR materials at the 20-30\% doping level.
The inefficiency of the off-diagonal disorder can be understood by looking
at the distributions of the amplitude of the hopping integral $t_{ij}$, 
$P(|t_{ij}|) = 2|t_{ij}|/t^2$. This distribution has small weight for small 
$|t_{ij}|$ which are most important for localization. The presence of the 
Berry phase, on the other
hand, weakens further the localization effect by breaking the time-reversal 
symmetry\cite{note3}. In fact the mobility edge without the Berry phase is 
located at $|E_c| \approx 3.3t$. The inefficiency of the pure DE model to 
localize electrons clearly points towards the necessity 
to include other effects, such as JT 
electron-phonon coupling, electron-electron interaction effects, and 
diagonal substitutional disorders. 

To investigate the effect of diagonal disorders, we have calculated
$\lambda_M$ for different values of W and E, similar to the W=0 case,
to determine the location of the mobility edge at fixed W. The
mobility edge trajectories in the W-E plane is shown as in Fig.~3.  
The shape of the mobility edge bears remarkable resemblance to the mobility
edge trajectory in the Anderson model with diagonal disorder alone\cite{bib21}.
The main difference is in the energy scale which is smaller in the
present system due to the smaller average value of the hopping integrals.
For the DE model, we obtain $\langle |t_{ij}|\rangle\ = \frac {2} {3}t$,
leading to an effective band edge\cite{note4} at $E_b =-4.0t$ at W=0. 
This reduction of band width also accounts for the smaller critical disorder
$W_c$ at the band center. The outward shift of the mobility edge for small
W is due to the increase of the effective band width with W. If E is
normalized with the effective band width instead of t,  
the region of extended states will always shrink with increasing disorder W. 
In the same figure, we have also plotted the equal-integrated-DOS lines.
Figure 3 shows that to achieve localization of 20-30\% of the electrons,
the presence of a substantial amount of diagonal disorders W, in the range of
10-12 in units of t, is required.

The ture mobility edge is controlled by the formation of localized polarons
in systems with strong electron-phonon couplings.  The presence of disorder 
changes the character of the polaron formation in three dimensions. In ordered 
systems, an abrupt change of the polaron state 
from nearly free type (large polaron) to self-trapped type (small 
polaron) occurs as the electron-phonon coupling reaches a critical 
value\cite{sumi}.
In disordered systems, however, localized polaron of intermediate sizes can 
form from the nonuniform extended electronic states above the mobility
edge even with moderate coupling\cite{cohen}. This transition occurs
at $\xi \approx 10\lambda^{-2/3}$, where $\lambda$ is the dimensionless 
electron-phonon coupling constant.  The true mobility edge, obtained by 
using the coherent length $\xi$ from the finite size scaling analysis
to be discussed below, is indicated in Fig.~3, for $\lambda=0.03$ and 1.
For large values of coupling constant, $\lambda >>1$, small polaron 
picture prevails. Figure 2 and 3  are the principal results of this work.

The presence of the Berry phase breaks time-reversal symmetry, and hence
the present model belongs to the unitary universality class.
The critical property around the MIT is investigated
using the one-parameter-scaling procedure\cite{bib20}.
We have been successful (Fig.~4) in placing all our data on the same
universal scaling curve,
$\lambda_m/M=f(\xi/M)$, where $\xi(W,E)$ is the scaling parameter
corresponding to the infinite size localization and correlation length
in the localized and metallic regimes respectively. Moreover, our scaling
data fall on the scaling function (shown as solid line in fig. 4) of
the standard Anderson model with diagonal disorder alone.
The critical exponent $\nu$ for the localization and correlation lengths,
$\xi \sim |E-E_c|^{-\nu}$, is found to be  $\nu = 1.0\pm0.2$, consistent
with the value for the standard Anderson model\cite{bib20}. Our result
further supports previous conclusions\cite{kramer2,schreiber} that the
critical property of the disorder-induced MIT is not modified by 
broken time-reversal symmetry in three dimensional systems.

In the Mn-oxides La$_{1-x}$A$_x$MnO$_3$, the potential fluctuation  
experienced by the $e_g$ electrons due to the La$^{3+}$ and A$^{2+}$ 
ion cores, if unscreened, would amount to $U\sim 0.6$ eV. This is 
equivalent to $W \approx 0.6 [12x(1-x)]^{1/2}$ eV = 0.95 eV at x=0.3.       
Using Gutzwiller approximation \cite{jun1} the band narrowing 
from e-e interactions at $V=20t$ is estimated to be around 0.5 
for $x=0.2$ and 0.6 for $x=0.3$, if the two orbitals are degenerate. 
Hence $W/t_{eff}$ can be as large as 13, if we assume t=0.15 eV. Screening
certainly will reduce this value, so our estimate is only meant to
stress the importance of disorder in MIT in Mn-oxides. We feel 
that disorder in CMR materials is not strong enough to induce 
localization by itself,  but its sizable presence can certainly modify small 
polaron formation picture\cite{millis-dm,jun1} for systems with 
strong but less than critical value of electron-phonon coupling.
This mechanism also has the advantage of not restricting to any particular
form of electron-phonon coupling, and thus may apply to a wide range 
of CMR materials described by the double exchange mechanism. 

In summary, we have investigated the charge 
localization properties in the Mn-oxides in the classical spin limit.
We find that a substantial amount of diagonal disorder is required to localize
20-30\% of electronic states. This suggests that large electron-phonon 
coupling and polaronic effect are necessary  to explain the MIT close to $T_c$. 

We would like to thank S.A. Trugman and H. R\"oder for helpful discussions.
Ames Laboratory is operated for the U.S. Department of Energy by Iowa
State University under Contract No. W-7405-Eng-82. This work was supported by 
the directorate for Energy Research, Office of Basic Energy Sciences. 
Work at Los Alamos is performed under the auspices of the U.S. DOE.

\begin{figure}
\centerline{
}
\caption{The ratio of the finite size localization length to the width,  
$\lambda_M/M$, as a function of M, for the double exchange model 
in the absence of any diagonal disorders. The mobility edge is located around 
$E_c =-3.56 t$. All states with $E<E_c$ are localized.}   
\end{figure}
\begin{figure}
\centerline{
}
\caption{Th integrated density of states of the double exchange 
model in the absence of any diagonal disorder, W = 0. The density
of states (in units of $t^{-1}$) is shown in the inset.
The results are obtained from exact 
diagonalization
of $10\times 10 \times 10$ clusters averaged over 100 random spin 
configurations.
}
\end{figure}
\begin{figure}
\centerline{
}
\caption{The mobility edge trajectory of the double exchange model with 
diagonal disorder W for electron-phonon coupling canstant 
$\lambda=$0, 0.03, and 1. 
The dashed lines show equal integrated density of states
lines at 10\%, 20\%, 30\%, and 40\% fractions. The density of states and
the mobility edge trajectory are symmetric around E=0 axis.
}
\end{figure}
\begin{figure}
\centerline{
}
\caption{The universal scaling function of the double exchange 
model for different W, E, and M. At the critical point, 
$\lambda_M/M \approx 0.6$.  
The solid line is the universal scaling curve
for the standard Anderson model with diagonal disorder alone.
}
\end{figure}


\begin{references}
\bibitem{note3}
This behavior should be distinguished from the localization effect due to purely
random phases in otherwise periodic systems, for example, in the random flux
problem that has been extensively investigated recently\cite{review}.
\bibitem{note4}
The true band edge, according to the Lifshitz argument, is located at
$\pm(6t+W/2)$.
\end{references}
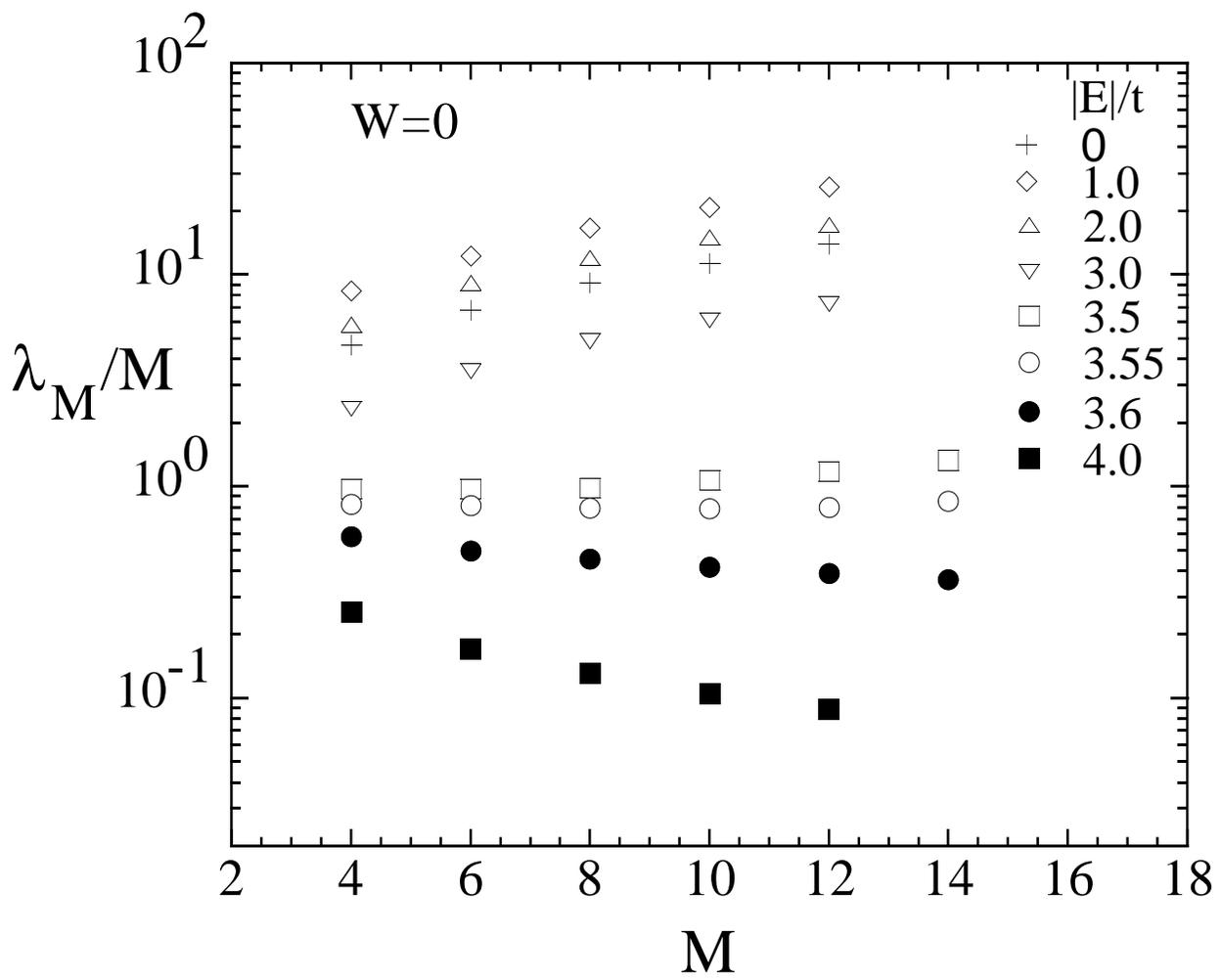
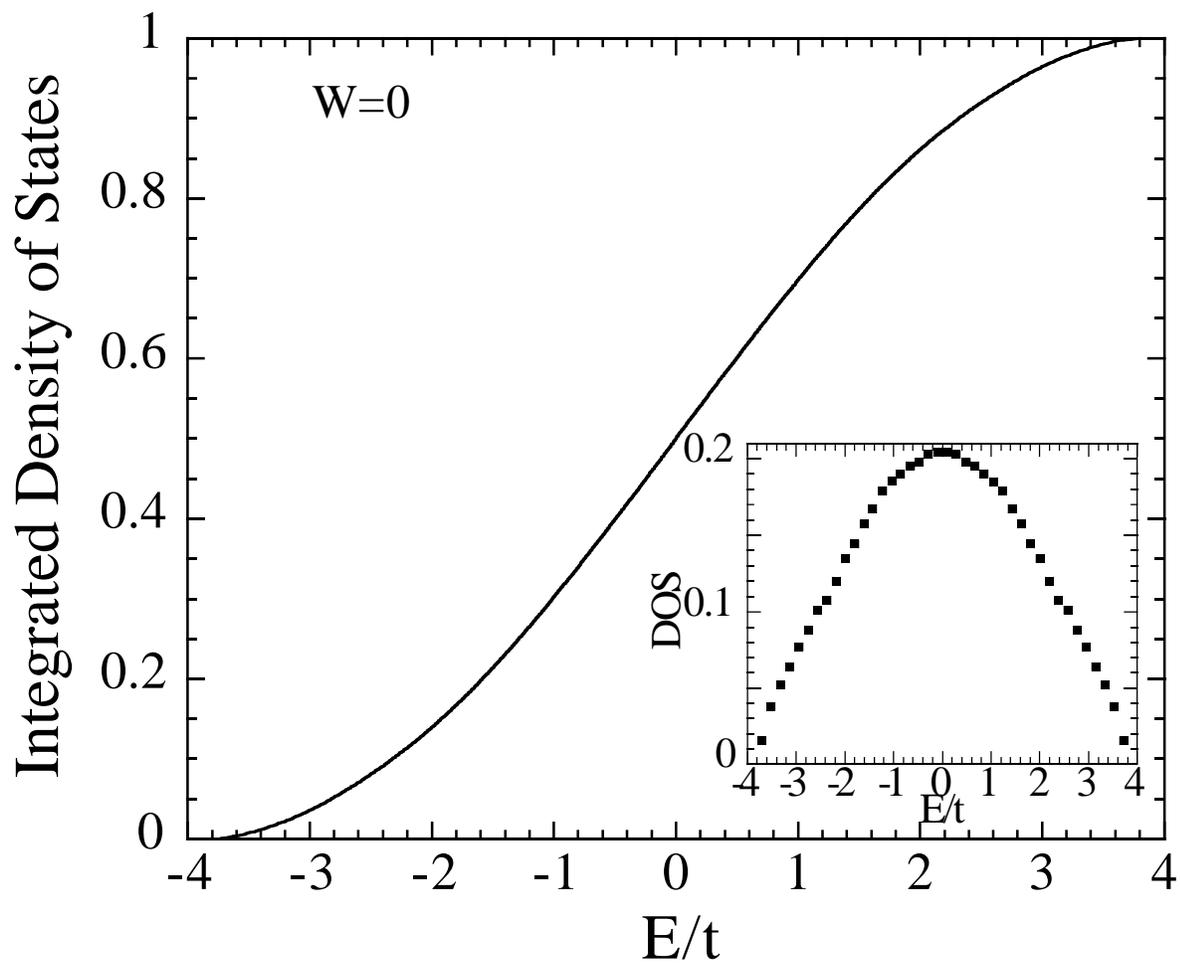
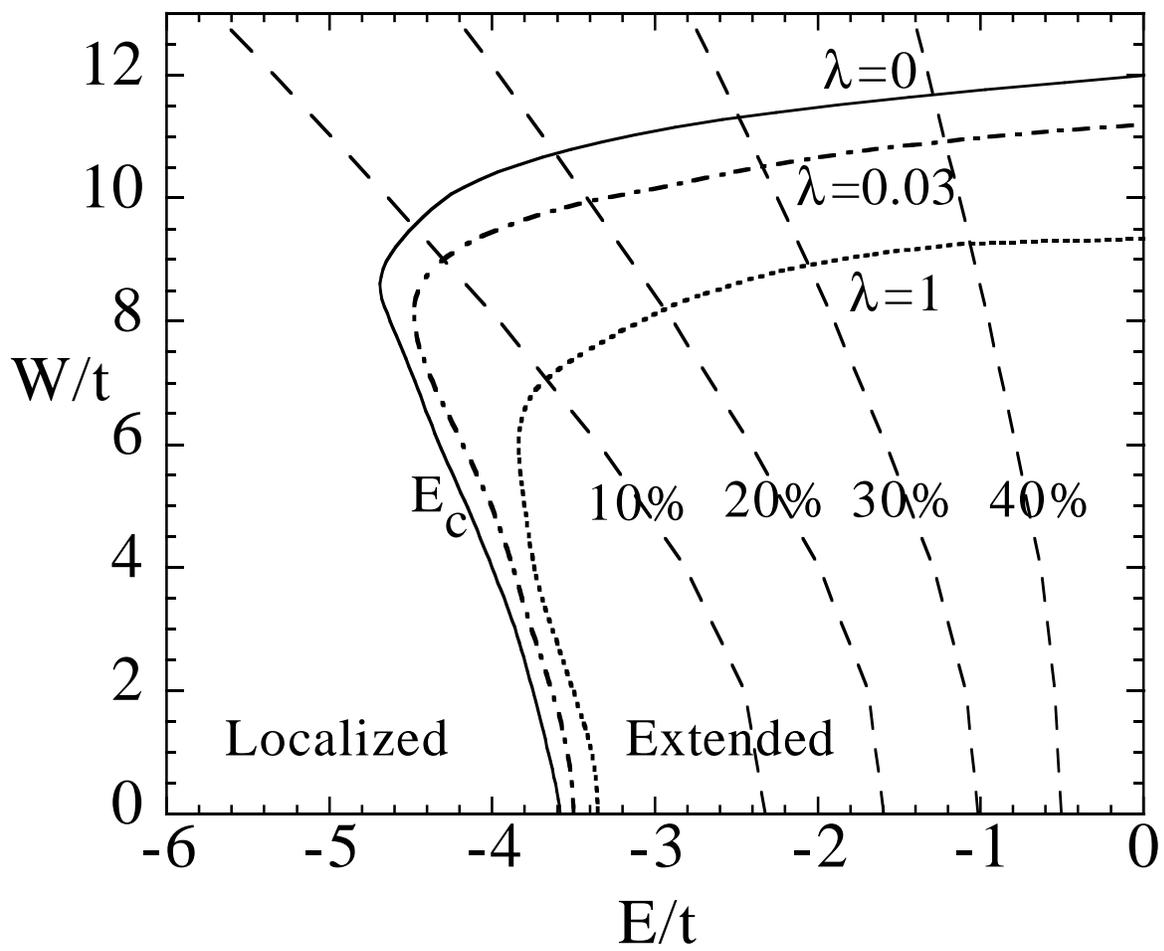
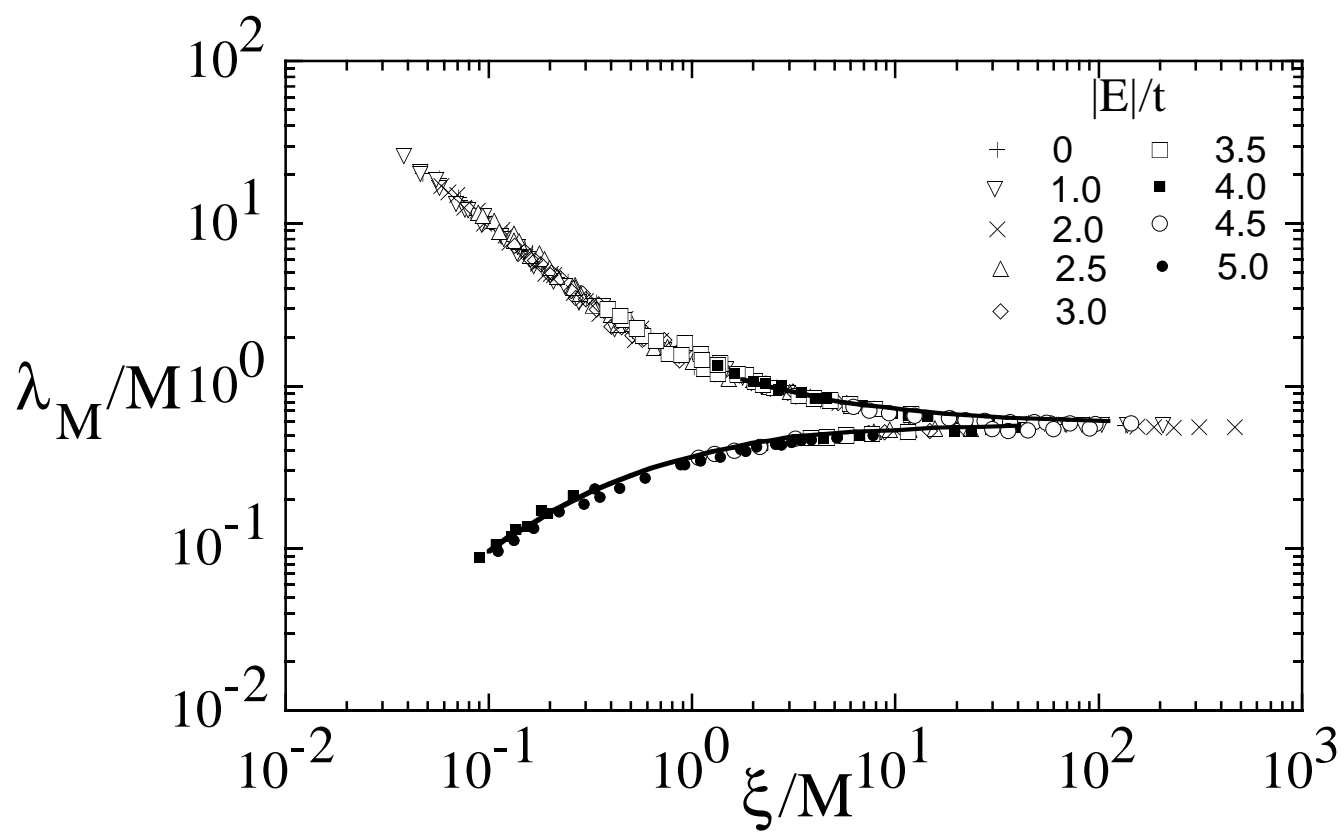
\end{document}